\documentclass[10pt,superscriptaddress]{revtex4-2}
\usepackage{amsmath}
\usepackage{amssymb}
\usepackage{bm}
\usepackage{epsfig}
\usepackage{graphicx}
\usepackage{color}
\usepackage[breaklinks,unicode=true,colorlinks=true,citecolor=blue,urlcolor=blue]{hyperref} 
\usepackage[utf8]{inputenc}
\usepackage[english]{babel}
\usepackage{xcolor}
\usepackage{array}
\usepackage[normalem]{ulem}
\usepackage{hyperbu}

\newcommand{\dd}[2]{\frac{\partial {#1}}{\partial {#2}}}

\begin{document}
\title{Efficient launch of shear phonons in photostrictive halide perovskites}
\author{D.~O.~Horiachyi}
\email[correspondence address: ]{dmytro.horiachyi@tu-dortmund.de}
\affiliation{Experimentelle Physik 2, Technische Universit\"at Dortmund, 44221 Dortmund, Germany}
\author{M.~O.~Nestoklon}
\author{I.~A.~Akimov}
\author{A.~V.~Trifonov}
\author{N.~V.~Siverin}
\author{N.~E.~Kopteva}
\author{A.~N.~Kosarev}
\author{D.~R.~Yakovlev}
\affiliation{Experimentelle Physik 2, Technische Universit\"at Dortmund, 44221 Dortmund, Germany}
\author{V.~E.~Gusev}
\affiliation{Laboratoire d’Acoustique de l’Universit\'e du Mans (LAUM), UMR CNRS 6613,\\ Institut d’Acoustique-Graduate School (IA-GS), Le Mans Universit\'e, Le Mans, France}
\author{M.~Fries}
\author{O.~Trukhina}
\author{V.~Dyakonov}
\affiliation{Experimental Physics 6 and W\"urzburg-Dresden Cluster of Excellence ct.qmat, Julius-Maximilians-Universit\"at W\"urzburg, 97070 W\"urzburg, Germany}
\author{M.~Bayer}
\affiliation{Experimentelle Physik 2, Technische Universit\"at Dortmund, 44221 Dortmund, Germany}
\affiliation{Research Center Future Energy Materials and Systems, Technische Universit\"at Dortmund, 44227 Dortmund, Germany}

\begin{abstract}
Optical generation of transverse coherent phonons by femtosecond light pulses is appealing for high-speed sub-THz active control of material properties. Lead-free double perovskite semiconductors, such as Cs$_2$AgBiBr$_6$, attract particular interest due to their cubic to tetragonal phase transition below room temperature and strong polaron effects from carrier-lattice coupling. Here, we reveal that the anisotropic photostriction in halide perovskites with tetragonal crystal structure represents an efficient non-thermal tool for generating transverse coherent phonons. In particular, we demonstrate that along with compressive strain, optical generation of photoexcited carriers leads to strong shear strain in Cs$_2$AgBiBr$_6$ below the phase transition temperature of 122~K. Using time-domain Brillouin spectroscopy, we observe coherent transverse and longitudinal acoustic phonons with comparable amplitudes in the tetragonal phase, while in the cubic phase only longitudinal phonons are generated. The polarization of the photoinduced transverse phonons is dictated by the projection of the $c$-axis on the surface plane, which leads to a prominent anisotropic polarization response in the detection. The generated strain pulses correspond to transverse acoustic soft eigenmodes with a strong temperature dependence of dispersion, which provides an additional degree of freedom for active hypersonic control.
\end{abstract}

\maketitle

\section{Introduction}

Terahertz (THz) and sub-THz coherent acoustic phonons show great potential for manipulations involving photons~\cite{Fainstein-2024,Brueggemann2012,Ortiz-21,Chern-2004,Matsuda-2002,Baldini-2019}, electrical currents~\cite{Sun-2004, Kent-2012} and fields~\cite{Ruello-2023}, magnons~\cite{Sherbakov-2010,Reppert-2020}, as well as plasmons~\cite{Temnov-2012}. Experiments have demonstrated that high frequency coherent phonons can be used for the generation of THz electromagnetic waves~\cite{Armstrong-2009, Rongione-2023}. Due to their nanometer wavelength, sub-THz phonons are applied for nano-imaging \cite{Audoin-2023}. Ultrafast spectroscopy of high-frequency coherent acoustic phonons employs the excitation and detection using short femtosecond laser pulses in pump-probe techniques, representing a powerful method to investigate the lattice dynamics in a wide spectrum of materials~\cite{Thomsen1986,Matsuda2015}. Typically, the generation of coherent phonons exploits the excitation of ultrafast stress associated with the thermoelastic coupling in metals \cite{Thomsen1986,Matsuda2015,Maris-1994,Wright-1994} or with the electronic deformation potential in semiconductors \cite{Thomsen1986,Young-2012, Lamitre-2012}. In most settings, the phonons are generated with polarization vector $\mathbf{U}$ parallel to the propagation direction and correspondingly to the phonon wave vector $\mathbf{Q}$, i.e. longitudinal acoustic (LA) phonons are provided. However, experiments with coherent phonons, whose polarization is perpendicular to $\mathbf{Q}$, i.e. transverse acoustic (TA, shear) phonons, have remained in high demand. Indeed, TA phonons possess a smaller sound velocity compared to LA phonons and, correspondingly, a smaller wavelength for the same frequency, of great advantage in imaging. Further, TA phonons posses two independent  polarization components and could be used to manipulate the spin properties of charge carriers by chiral acoustic waves~\cite{Korenev-2016}.

Generation of TA phonons requires a shear perturbation of the crystalline lattice. The most common method to induce shear strain in a material is to excite a crystalline surface of low symmetry by an ultrafast optical pulse ~\cite{Gusev-2007,Matsuda2008,Kent-2017}. In this case, even isotropic compressive stress results in both compressive and shear dynamical strain and leads to the generation of quasi-LA and quasi-TA phonons with wave vectors $\mathbf{Q}$ perpendicular to the optically excited surface. Also, the piezoelectric mechanism of phonon generation may have significant importance~\cite{Sun-2012,RUELLO201521}. Usually, the amplitude of the generated TA phonons is much smaller than that of the LA phonons. Only recently, a few studies have reported amplitudes of the generated TA phonons comparable with those of the LA phonons~ \cite{Lejman-2014, Guo-2017, Mante-2018}, but the mechanisms of efficient generation of shear acoustic waves are still under debate. Here, material systems such as the perovskites exhibiting a variety of structural phase transitions, provide a rich testbed for exploring new scenarios of TA phonon generation. Furthermore, semiconductor perovskites are particularly interesting for exploiting photoinduced mechanisms as the interaction of optically excited carriers with the crystal are dominant here.

In recent years, the interest in perovskite semiconductors has grown rapidly due to their success in photovoltaic applications~\cite{Roadmap-2021}. Inorganic lead-free double perovskites are particularly relevant here as non-toxic and stable material platform, but they so far suffer from inefficient charge generation and transport, for which the responsible mechanisms are yet not well understood~\cite{Wu-2021,Tress-2022}. Among them are complex phenomena associated with structural phase transitions (cubic to tetragonal)~\cite{Schade2019} and strong electron-phonon interactions (polaron effects)~\cite{Wu-2021, Baranowski-2019}, which can significantly influence the mobility of charge carriers~\cite{Herz-THz-2021}. In this respect, double perovskite semiconductors are particularly exciting for ultrafast acoustics. First, the elastic constants are about three times smaller as compared to conventional semiconductors such as GaAs~\cite{Even-2016}. A soft crystal lattice allows one to achieve a significantly larger deformation for the same magnitude of stress. Second, despite the absence of the piezoelectric contribution leading to a strong renormalization of the photoinduced stress in ferroelectric perovskites \cite{Paillard2017}, in lead halide perovskites an analogous renormalization due to structural phase transition is present~\cite{Paillard2023}. Moreover, the formation of exciton polarons, which distort the lattice after pulsed optical excitation, is not captured by the ``deformation potential'' mechanism of coherent phonon generation. Finally, the ease of their synthesis as well as the rapid development of growth technology of perovskite semiconductors underline the potential for deposition of thin films on top of other materials which could subsequently lead to the realization of optically driven phononic transducers \cite{Du2021}.

Here, we reveal and elaborate a new mechanism for efficient optical generation of shear strain in perovskite semiconductors via the anisotropic photostriction of the tetragonal crystal lattice. Coherent TA phonons with amplitude comparable to the LA modes are observed in the tetragonal phase of a Cs$_2$AgBiBr$_6$ crystal using time-domain Brillouin spectroscopy after resonant excitation of excitons by femtosecond laser pulses. We demonstrate that only one of the two TA modes is excited via the anisotropic photostriction whose polarization direction is given by the projection of the $c$-axis on the sample surface. This mode corresponds to a soft TA phonon with a sharp temperature dependence of the sound velocity in the vicinity of the structural phase transition occuring at a temperature of about 122~K, as confirmed by both time-resolved and continuous wave ($cw$) Brillouin light scattering (BLS). Our results show that perovskite semiconductors and in particular lead-free double perovskites represent an attractive system for implementing active hypersonic devices where frequency and polarization of the acoustic phonons can be tuned across a wide range.

\section{Ultrafast optical response in \texorpdfstring{$\mathbf{Cs_2AgBiBr_6}$}{Cs2AgBiBr6}}
\label{sec:TDBLS}

\begin{figure}
	\includegraphics[width = 0.5\columnwidth]{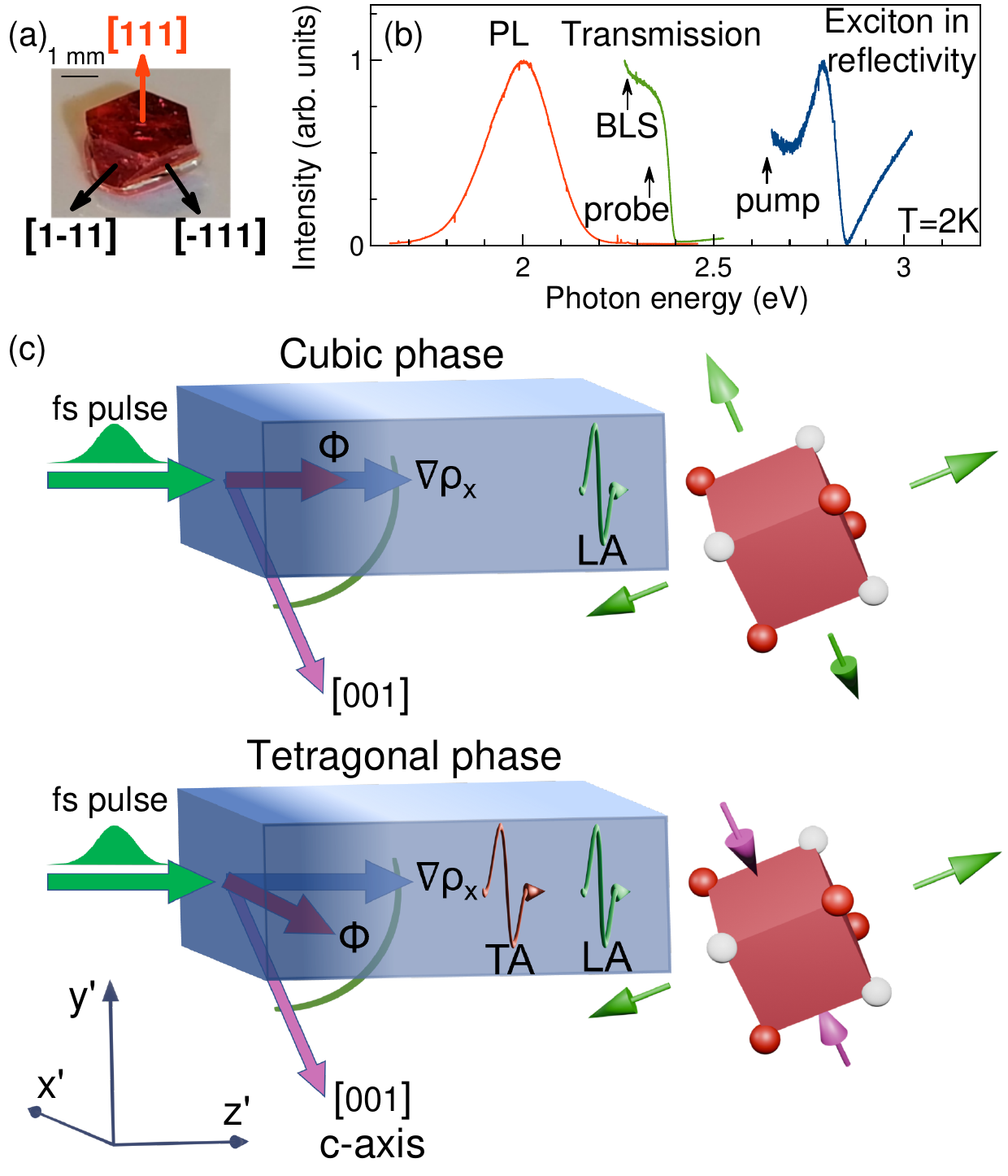}
	\caption{{\bf Ultrafast generation of strain pulses.} 
(a) The single Cs$_2$AgBiBr$_6$ crystal under study. Arrows point along the main crystallographic directions of the available facets. 
(b) Reflectivity (blue), transmission (green), and photoluminescence (PL, red) measured at $T=2$~K. The reflectivity spectrum shows the exciton resonance at 2.82~eV. The transmission spectrum is measured in the vicinity of the absorption edge. Vertical arrows indicate the photon energies in the time-resolved pump-probe and continuous wave ($cw$) Brillouin light scattering (BLS) experiments.  (c) Schematic presentation of the excitation of a strain pulses by a fs optical pump pulse along the $\mathbf{z}'$ direction. The color gradient indicates the surface layer containing the photoexcited carriers. In the cubic phase, the photogenerated stress driving force $\mathbf{\Phi}\|\mathbf{z}'$ results in compressive strain. Only a LA phonon pulse is launched in this case, as shown by the green wiggly arrow. In the tetragonal phase, the expansion coefficient along the [001] direction (c-axis) is different from that along the [100] and [010] directions. Then the driving force $\mathbf{\Phi}$  acquires an in-plane component (along $y'$) and leads to shear strain. If the expansion coefficients have different signs (see green and magenta arrows), the shear strain is significant, giving rise to TA and LA pulses with comparable amplitudes (red and green wiggly arrows).
} 
	\label{fig:generalInfo}
\end{figure}

Single crystals of Cs$_2$AgBiBr$_6$ were grown from a mixture of CsBr, AgBr and BiBr in a 2:1:1 molar ratio in 48\%~wt HBr~\cite{Adam-2016, Armer-2021}. The reaction flask was heated to 110$^\circ$C for complete dissolution of the metal salts, and cooled down at a rate of 1$^\circ$C/h to room temperature (see Methods). The simple cubic crystal structure in ambient conditions was proven by powder X-ray diffraction (see Supplementary Information section~\ref{sec:si:XRD}). All of the crystal facets are associated with (111) or equivalent crystallographic planes as shown in Fig.~\ref{fig:generalInfo}(a). Note that in the tetragonal phase these planes have the Miller indices (011) and equivalent.

\begin{figure*}[ht]
\includegraphics[width=0.8\linewidth]{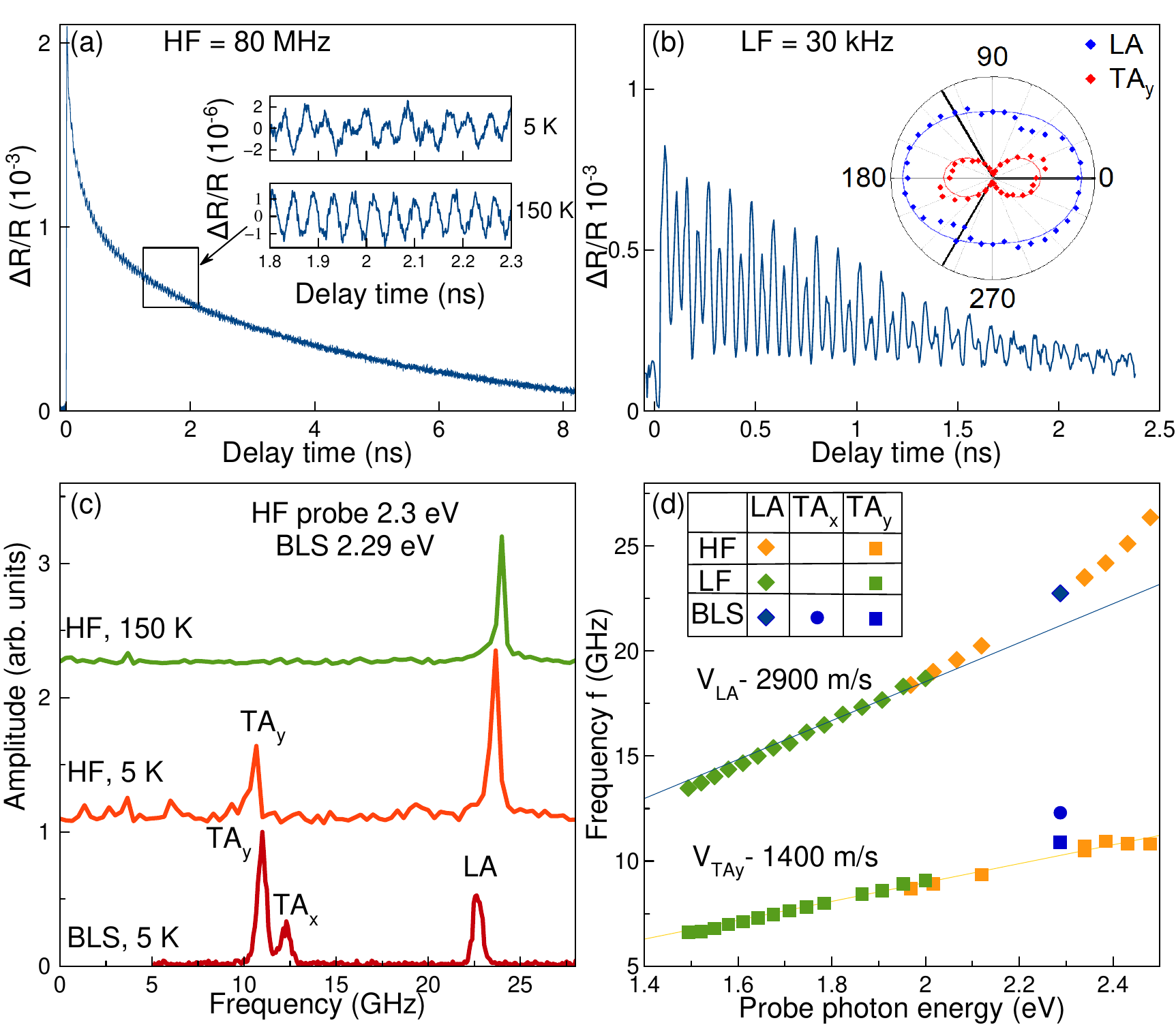}
\caption{
{\bf Dynamics of pump-induced reflectivity modulation.} (a) and (b) Examples of pump-probe transients of differential reflectivity $\Delta R/R$ from the Cs$_{2}$AgBiBr$_{6}$ crystal at $T=5$~K, i.e. in the tetragonal phase, measured along the crystal direction [011] ([111] in the cubic phase) in HF and LF experiments with repetition frequencies of 80~MHz and 30~kHz, respectively. In (a) the photon energies of the pump $h\nu_{\rm pump}$ and probe $h\nu_{\rm probe}$ pulses are set to 2.638 and 2.339~eV, respectively. The insets show fast oscillatory signals below and above the structural phase transitions, taken at $T=5$~K and 150~K. In (b) $h\nu_{\rm pump}=2.818$~eV and $h\nu_{\rm probe}=2$~eV. The inset shows the polar plot of the amplitudes of the fast oscillatory components with frequencies $f_{\rm LA}=18.6$~GHz and $f_{\rm TA_{\rm y}}=9$~GHz (blue and red diamonds, respectively) as function of the direction $\theta$ of the probe pulse polarization relative to the $y'$-axis. The solid black lines show possible projections of the directions of the c-axis on the crystal surface. The solid red and blue lines are fits using the function $A-B\sin^2(\theta)$. (c) The top and middle curves correspond to fast Fourier transform (FFT) spectra of the pump-probe signals in (a), measured at 5~K (red) showing two peaks in the tetragonal phase, and measured at 150~K (green) with only one peak in the cubic phase. The lower spectrum corresponds to {\it cw} Brillouin light scattering (BLS) measured in the back reflection geometry at the photon energy $h\nu_{\rm BLS}=2.287$~eV for $T=5$~K. (d) Frequencies of the FFT peaks of the coherent longitudinal LA (diamonds), transverse acoustic TA$_{\rm x}$ (circles) and TA$_{\rm y}$ (squares) phonons as function of the probe photon energy. The $cw$ BLS data are shown with the blue symbols, the LF data with the green symbols, and the HF data with the yellow symbols. Linear fits through zero frequency are presented by the solid blue and yellow lines.}
\label{fig:PumpProbeData}
\end{figure*}

The optical properties of the investigated sample are summarized in Fig.~\ref{fig:generalInfo}(b). The low temperature reflectivity spectrum (blue line) shows a strong feature centered at about 2.822~eV, which is associated with the exciton resonance corresponding to the direct interband optical transition at the $\Gamma$-point~\cite{Baranowski-2019}. The green and red lines in Fig.~\ref{fig:generalInfo}(b) show the transmission and photoluminescence (PL) spectra with a sharp absorption edge at about 2.4~eV and a broad PL peak at about 2.0~eV, respectively, in agreement with previous studies~\cite{Baranowski-2019,Herz-THz-2021}. Note that these values cannot be directly attributed to the indirect band gap due to the strong electron-phonon coupling with a Huang-Rhys factor in the order of 10~\cite{Baranowski-2019}.

For investigation of the optically generated strain pulses, we use the well established method of time-domain Brillouin spectroscopy~\cite{RUELLO201521}. As shown in Fig.~\ref{fig:generalInfo}(c), the pump pulse with a photon energy above the absorption edge generates photoexcited carriers in the thin surface area of the crystal, which leads to the formation of a stress driving force $\mathbf{\Phi}$ and to subsequent propagation of a strain pulse away from the surface along the $\mathbf{z}'$ direction, i.e. along the gradient of the photoexcited carrier density $\nabla\rho_X$. In isotropic or cubic crystals with high symmetry crystallographic directions, $\mathbf{\Phi}\|\mathbf{z}'$ so that only longitudinal strain pulses are generated.  However, as will be shown below, anisotropic expansion and, in particular, simultaneous expansion and compression along different directions can also induce a force with a direction different from $\mathbf{z}'$, which subsequently leads to the generation of shear strain pulses (TA pulse in Fig.~\ref{fig:generalInfo}(c)). 

Time-resolved pump-probe measurements were performed using two different laser repetition frequencies of 80~MHz and 30~kHz, which are below referred to as high (HF) and low (LF) frequency, respectively. In the HF case the energy fluence of the pump laser $\Psi_{\rm pump}$ is weak corresponding to about 30~$\mu$J/cm$^2$, while for the LF measurements $\Psi_{\rm pump} \sim$ 700~$\mu$J/cm$^2$ is significantly stronger. The duration of the laser pulses is about 150~fs (see Methods).   

Recorded pump-probe results are summarized in Fig.~\ref{fig:PumpProbeData}, namely examples of differential reflectivity $\Delta R/R$ transients at $T=5$~K  temperature with HF and LF excitation in Figs.~\ref{fig:PumpProbeData}(a) and~\ref{fig:PumpProbeData}(b), respectively.  In all cases the signal can be described by fast oscillations superimposed on a slowly varying decaying transient. The latter originates mainly from carrier relaxation after pulsed photoexcitation which depends on the energy of the laser pulse, i.e. varies for HF or LF excitation. For HF excitation we observe a biexponential decay with characteristic times of 0.3~ns and 4~ns, while for the LF excitation a single exponential decay with about 2~ns decay time is found. We note that the relative magnitude of the oscillatory signal is much more pronounced for LF excitation, which is related to the larger energy in an excitation pulse. In what follows, we focus on the fast oscillatory signal to be analyzed after subtraction of the slowly varying exponential signal, as shown exemplary in the inset of Fig.~\ref{fig:PumpProbeData}(a).

Surprisingly, at the low temperature of $T=5$~K, when the crystal is in the tetragonal phase, the signal comprises two frequencies $f_i$ of about 10 and 23~GHz, see Fig.~\ref{fig:PumpProbeData}(c). At temperatures above $T_c\approx122$~K, the crystal structure transforms into the cubic phase~\cite{Schade2019}. Here, only the component with the higher frequency persists, as shown in the insets of Fig.~\ref{fig:PumpProbeData}(a) and Fig.~\ref{fig:PumpProbeData}(c) for the temperature of 150~K. The normalized fast Fourier transform (FFT) spectra of the oscillatory signals for HF excitation shown in Fig.~\ref{fig:PumpProbeData}(a) are presented in Fig.~\ref{fig:PumpProbeData}(c). The frequencies are independent of the photon energy of the pump pulses, but increase linearly with  increase of the probe photon energy $h\nu_{\rm probe}$, see Fig.~\ref{fig:PumpProbeData}(d).

This behavior is observed in time-domain Brillouin scattering on acoustic phonons, i.e. when the probe pulse is scattered from the pump-induced strain pulse propagating away from the surface with the speed of sound $V_i$~\cite{RUELLO201521}. In this case, the oscillation frequency is determined as $f_i = 2n_r \nu_{\rm probe}V_i/c$, where $c$ is the speed of light and $n_r$ is the refractive index in Cs$_2$AgBiBr$_6$. Taking the refractive index at room temperature $n_r=2$ from Ref.~\cite{Huygen2021}, we fit the data in Fig.~\ref{fig:PumpProbeData}(d) with a linear function and estimate the speed of sound for the LA and one of the TA modes as $V_{\rm LA}=2900$~m/s and $V_{\rm TA_y}=1400$~m/s at $T=5$~K, which are close to the data evaluated by $cw$ Brillouin light scattering~\cite{Lun2022}. 

We also performed $cw$ BLS measurements on the same sample in reflection geometry along the direction normal to the surface, using a $cw$ single frequency laser with photon energy $h\nu_{\rm BLS} = 2.287$~eV and a spectral width of about 40 neV. The spectrum taken at $T=5$~K in the tetragonal phase is shown in Fig.~\ref{fig:PumpProbeData}(c). It comprises three lines with frequencies of $f_{\rm LA} = 22.8$~GHz, $f_{\rm TA_x} = 12.3$~GHz, and $f_{\rm TA_y} = 10.9$~GHz corresponding to the longitudinal and the two transverse acoustical phonons. For more details about the $cw$ BLS data, see the Supplementary Information section~\ref{sec:si:BLS}. The frequencies of the peaks are also in agreement with the photon energy dependence of the pump-probe data in Fig.~\ref{fig:PumpProbeData}(d) (see blue symbols). Thus, we attribute the higher and lower frequencies in the pump-probe transients to the LA and TA$_{\rm y}$ phonons, respectively. The slight deviation from the linear dependence at photon energies close to the bandgap ($h\nu_{\rm probe}\sim 2-2.4$~eV) is probably related to changes in the refractive index (see Fig.~\ref{fig:PumpProbeData}(d)).

In summary, we highlight that high-frequency coherent TA phonons are detected in the tetragonal phase, along with LA phonons, in time-domain Brillouin spectroscopy. In contrast, in the cubic phase, only LA phonons are generated. Interestingly, only one of the two TA phonons is detected, with an amplitude comparable to that of the LA signal. The polarization dependence of the TA signal is strongly anisotropic, with the peak amplitude obtained for the probe polarization having maximum projection on the c-axis, see the inset in Fig.~\ref{fig:PumpProbeData}(b). The measured polarization angular dependence follows the form $1-0.29\sin^2(\theta)$ for the LA and $0.55-0.5\sin^2(\theta)$ for the TA acoustic phonons. Note that in our experiments, the c-axis projection on the sample surface was randomly directed along one of the three directions of the facet edges after every cooling cycle from the cubic to the tetragonal crystallographic phase. That is, in the sample coordinates, the angle $\theta$ is counted from one of these directions, which are equivalent in the cubic phase. In what follows we evaluate the origin of the strong TA signal in the pump-probe transients.


\section{Optical generation of shear strain in tetragonal phase}
For analysis of the optical generation of strain pulses one needs to consider first the eigenmodes of the acoustic waves and the basic properties of their polarization. Further, we need to consider the mechanisms of conversion of the heat and carriers generated by the incident optical pulse to the deformation pulse propagating through the crystal. The eigenmodes of the acoustic waves can be represented as plane waves with polarization vectors ${\bf U}^s$ given by 
\begin{equation}\label{eq:eigen_ph}
  \left( \rho V_s^2 \delta_{il} - c_{ijkl} Q_j Q_k \right) U_l^s =0\,.
\end{equation}
Here $\rho$ is the mass density of the material, $c_{ijkl}$ are the elastic constants of the material given in Section~\ref{sec:si:tensors} of the Supplementary Information, ${\bf Q}$ is the direction of the phonon wave vector, $V_s$ is the speed of sound for the $s$-th phonon mode, $\delta_{il}$ is the Kronecker delta. The solution of Eq.~\eqref{eq:eigen_ph} is discussed in more detail in the Supplementary Information, Section~\ref{sec:si:phonons}. In a cubic system, from Eq.~\eqref{eq:eigen_ph} it follows that the LA phonons are polarized along $\mathbf{z}'$, while the two TA phonon modes are degenerate and their polarization vector ${\bf U}^s$ may be chosen along $\mathbf{x}'$ and $\mathbf{y}'$:
\begin{equation}\label{eq:coords}
  \mathbf{x}'||[1\bar10]\,,\,\,\,
  \mathbf{y}'||[11\bar2]\,,\,\,\,
  \mathbf{z}'||[111]\,.
\end{equation}
To understand the origin of the photoinduced shear strain in the tetragonal phase (note that in the tetragonal phase, the Miller indices of these directions \eqref{eq:coords} are $[100]$, $[01\bar1]$, $[011]$), one should distinguish two effects: (i) the change of velocities, which does not change the phonon polarization, and (ii) the change of polarization vector of phonons with respect to the axis frame. Calculation of the elastic constants using a density functional theory (DFT) approach (see Supplementary Information~\ref{sec:si:tensors} for details) renders the following information about polarization mode mixing: (a) for the LA mode, a $\sim 0.8\%$ admixture of the TA$_1$ mode and a $\sim 6.5\%$ admixture of the TA$_2$ mode; (b) for the TA$_1$ mode, admixtures of $\sim 5\%$ of the TA$_2$ mode and $\sim 0.4\%$ of the LA mode; and (c) for the TA$_2$ mode, admixtures of $\sim 5\%$ of the TA$_1$ mode and $\sim 6.5\%$ of the LA mode. This result assumes that, while the change of sound velocities (which in particular splits the TA mode frequencies) is large, its effect on the polarization directions may be calculated perturbatively (see the Supplementary Information~\ref{sec:si:phonons} for the eigenmodes in the tetragonal phase). As a result, in the tetragonal phase, the TA$_1$, TA$_2$, and LA phonons are predominantly polarized along the axes $\mathbf{x}'$, $\mathbf{y}'$, $\mathbf{z}'$, respectively. This means that the large amplitude of the shear strain observed in the experiment cannot be explained by mixing between the TA and LA modes, but requires a different mechanism.

The amplitude $A_s$ of the generated strain pulse after laser excitation is proportional to the product of the polarization of the $s$-th phonon mode ${\bf U}^s$ and the ``driving force'' $\mathbf{\Phi}$ whose components are given by (see details in Sec.~\ref{sec:si:strain_wave}) 
\begin{equation}
  \Phi_i=\dd{\Sigma_{ij}}{x_j}\,, 
\end{equation}
where $\Sigma$ is the stress tensor induced by the pump laser pulse. As explained above, the direction of ${\bf U}^s$ only slightly deviates from $\mathbf{x}'$ (TA$_1$), $\mathbf{y}'$ (TA$_2$), and $\mathbf{z}'$ (LA). This means that the amplitude of the generated phonons is proportional to the components of the driving force $\mathbf{\Phi}$.

The stress tensor contains two main contributions. The first one is given by the linear expansion of the lattice induced by heat~\cite{Thomsen1986} which is proportional to the increase of the lattice temperature $\delta T$:
\begin{equation}\label{eq:sigma_dT}
  \Sigma^{T}_{ij} = - 3 \delta_{ij} B \beta_i \delta T(z')\,,
\end{equation}
where $\beta$ is the linear expansion temperature coefficient and $B$ is the bulk modulus. Another one is proportional to the density of the photoexcited carriers $ \rho_X $:
\begin{equation}\label{eq:sigma_nX}
  \Sigma^{n}_{ij} = \delta_{ij} B \alpha_{i} \rho_X (z')\,.
\end{equation}
Here, $\alpha_i$ is the photostriction coefficient, i.e. the change of lattice constant by the photogenerated electron-hole pairs. In the literature, the stress due to the presence of carriers is called either ``deformation potential'' or ``photostriction''. We use the second term to highlight that the deformation potential is not the only mechanism contributing to the phenomenon described by Eq.~\eqref{eq:sigma_nX}.  As shown recently for lead halide perovskites in Ref.~\cite{Paillard2023},  the deformation potential alone is insufficient to describe the effect of the charge concentration on stress, not even qualitatively.

In GaAs the strain caused by lattice heating as described by Eq. \eqref{eq:sigma_dT} is estimated to be $\sim 3\%$ from the contribution given by Eq.~\eqref{eq:sigma_nX}~\cite{Matsuda2005}. In MAPbI$_3$, the photostriction coefficient is estimated \cite{Zhou2016} to be at least two orders of magnitude larger than in GaAs. Therefore, it appears to be consequent to expect that this contribution is the dominating one in (double) perovskites. 

Referencing to the surface which is the plane with normal $-{\bf n}$, both the temperature increase and the density of the photogenerated carriers are given by exponential functions:
\begin{equation}
  \delta T({\bf r})\,,\;\rho_X({\bf r}) \propto {\exp(-{\bf n}\cdot{\bf r}/\zeta)}\,,
\end{equation}
where $\zeta$ is the absorption length with $\bf r$ being the coordinate. In case of isotropic thermal expansion and photostriction the driving force is directed along the surface normal $\bm{\Phi} \| {\bf n}$ and only LA phonons can be generated.  
However, for materials of low symmetry, the linear expansion coefficient may be different for different crystallographic directions~\cite{ITCBD}. The thermal coefficients for the linear expansion in Cs$_2$AgBiBr$_6$ are almost equal in magnitude, but have different signs for the directions perpendicular to and along the $[001]$ crystal axis \cite{Schade2019}. This behavior is also typical for tetragonal phase perovskites: in $\beta$-MAPbI$_{3}$ (space group \#140, $I4/mcm$), $\beta_{\perp c} \approx -\beta_{||c}$ \cite{Jacobsson2015}.

To our knowledge, neither measurements nor calculations of the photostriction coefficients for Cs$_2$AgBiBr$_6$ are available. However, calculations for $\beta$-CsPbI$_3$ show that $\alpha$ has different signs perpendicular to and along the $c$-axis ($\alpha_{\perp c}/\alpha_{||c} \approx -0.6$)~\cite{Paillard2023}, similar to the temperature expansion coefficient.
In CsPbI$_3$, this large anisotropy has been qualitatively explained by accounting for the cubic-to-tetragonal phase transition~\cite{Paillard2017,Paillard2023}. Recently, it also has been demonstrated that one of the main contributions to the electron-phonon interaction in CsPbI$_3$ is the interaction with the LO phonon modes associated with phase transitions \cite{Hoffman2023}. Therefore, we expect that a similarly anisotropic photostriction is present in double perovskites showing cubic to tetragonal phase transition and strong polaron effects.

\begin{figure}[t]
 \centering{\includegraphics[width=0.3\linewidth]{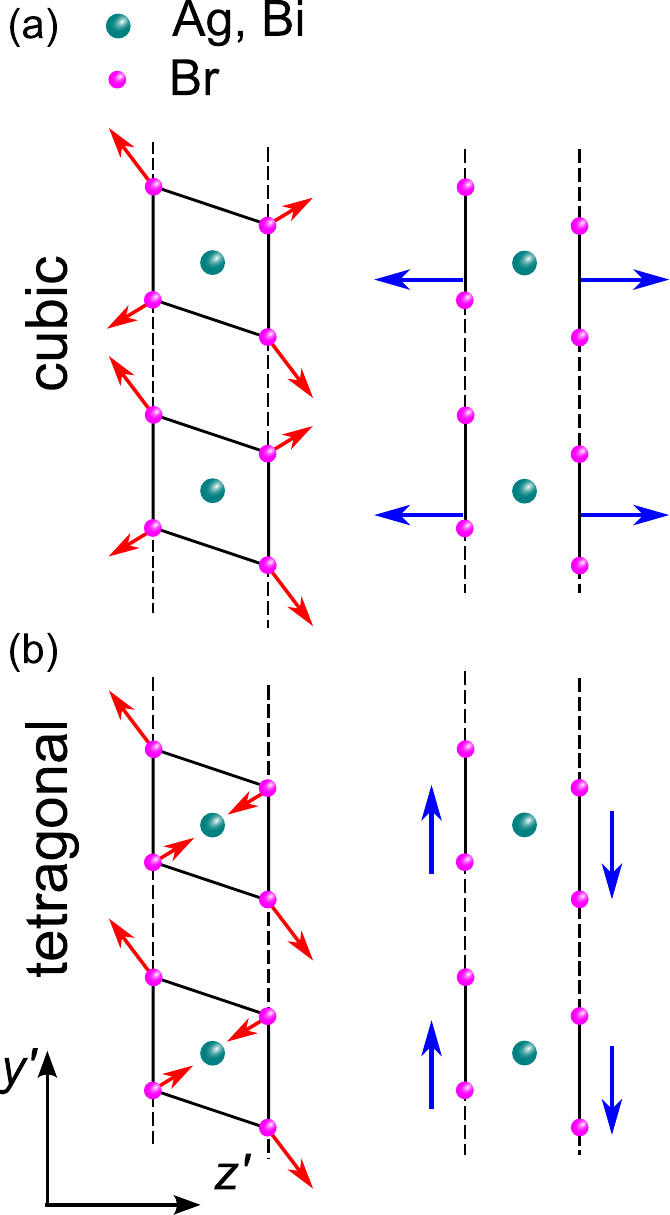}}
\caption{{\bf Shear strain generation.} Sketch illustrating the occurrence of a transverse component of the driving force in the tetragonal phase. The colored spheres show the positions of the atoms, the red arrows indicate the direction of the force acting on the individual atoms and the blue arrows show the average stress force acting in the surface plane. The red arrows with shorter lengths result from the projection of the stress vector into the $y'z'$ plane. They have to be taken into account twice due to the two atoms in the cell along the $\mathbf{x}'$ axis.}
 \label{fig:Sketch}
\end{figure}

As a result of the anisotropy mentioned above, the driving force acquires an in-plane component if the c-axis is neither parallel to the sample surface nor normal to it. Note that this does not require excitation of a surface of low symmetry. For example, in our case, this condition is fulfilled for optical excitation of the $x'y'$ facet. The deviation of $\alpha_{\perp c}$ from $\alpha_{||c}$ results in rotation of the driving force $\mathbf{\Phi}$ in the $y'z'$ plane, see Fig.~\ref{fig:generalInfo}(c) and Fig.~\ref{fig:Sketch}. The $y'$ component of $\mathbf{\Phi}$ generates TA$_2$ phonons. For $\alpha_{\perp c} \simeq -\alpha_{||c}$, the amplitude of the generated TA$_2$ phonons is $2\sqrt2$ times larger than the amplitude of the generated LA phonons. Note that TA$_1$ phonons could generally also be generated, because the elastic constant $c_{16}$ is non-zero in double perovskites. This is in contrast to the tetragonal phase in lead halide perovskites, where this process is forbidden by symmetry ($c_{16}=0$). Nevertheless, according to the experiment on Cs$_2$AgBiBr$_6$, only one of the TA phonon modes is excited, which is explained by the small magnitude of $c_{16}$. Hence, the transverse modes labeled TA$_{\rm x}$ and TA$_{\rm y}$ in the experimental data of Fig.~\ref{fig:PumpProbeData} correspond to transverse phonons polarized along the $\mathbf{x}'$ (TA$_1~\equiv {\rm TA_x}$) and $\mathbf{y}'$ (TA$_2 ~\equiv {\rm TA_y}$) axes, respectively. Based on this analysis, we can identify that the TA$_{\rm y}$ mode has a lower frequency than the TA$_{\rm x}$ mode. Indeed, as follows from Fig.~\ref{fig:PumpProbeData}(d), the frequency of the low energy TA phonon in the $cw$ BLS spectra (blue square) coincides with the only mode (TA$_{\rm y}$) excited in the pump-probe experiment (yellow squares). 

The detected signal arises from the reflection of the probe pulse at the strain wave. Its strength is given by the strain-induced variation of the dielectric constant $\Delta\hat{\varepsilon}$ \cite{Matsuda2008,Subbaswamy1978} which in turn is proportional to the strain tensor $\hat{\epsilon}$~\cite{Nelson70,Anastassakis74,Vacher72}. 
As shown in Sec.~\ref{sec:si:phonons} of the Supplementary Information, the strain tensor for different modes is proportional not only to the amplitude $\sim {\bm{\Phi}}\cdot {\bf U}^s$, but also to $\epsilon^s_{ij} = \frac12 (U_i^s Q_j + U_i^s Q_i)$.

For the general photoelastic tensor respecting the tetragonal symmetry, all relevant components of the dielectric constants may be non-zero, leading to light scattering on all possible phonons with a complex polarization dependence, see the Supplementary Information Sec.~\ref{sec:si:detect} for details. In the opposite limiting case of a fully isotropic material, the probe signal would be reflected only from the LA strain pulse independent of its polarization and would be proportional to $p_{1122}$. None of these scenarios appears in our experiment. We remind that it is necessary to account for both the LA and TA$_{\rm y}$ strain pulses, which are predominantly excited by the pump laser pulses. From the inset of Fig.~\ref{fig:PumpProbeData}(b), it follows that scattering of the probe on LA phonons shows a weak polarization dependence (blue ellipsoid), while for the TA$_{\rm y}$ phonons, scattering takes place predominantly for light polarized along the direction of the $c$-axis (two-lobed red rosette). According to the theoretical analysis in Sec.~\ref{sec:si:detect} of the Supplementary Information, this means that the opto-elastic tensor is almost isotropic in the plane normal to the $c$-axis, and $p_{1122}\approx p_{1133}$. We emphasize that the strong polarization dependence for the scattered probe beam provides a simple procedure for determining the direction of the $c$-axis in the tetragonal crystal phase by optical methods.


\section{Temperature dependence: tetragonal vs cubic phase}
\label{sec:T-dep}

\begin{figure}[b]
\includegraphics[width=0.5\linewidth]{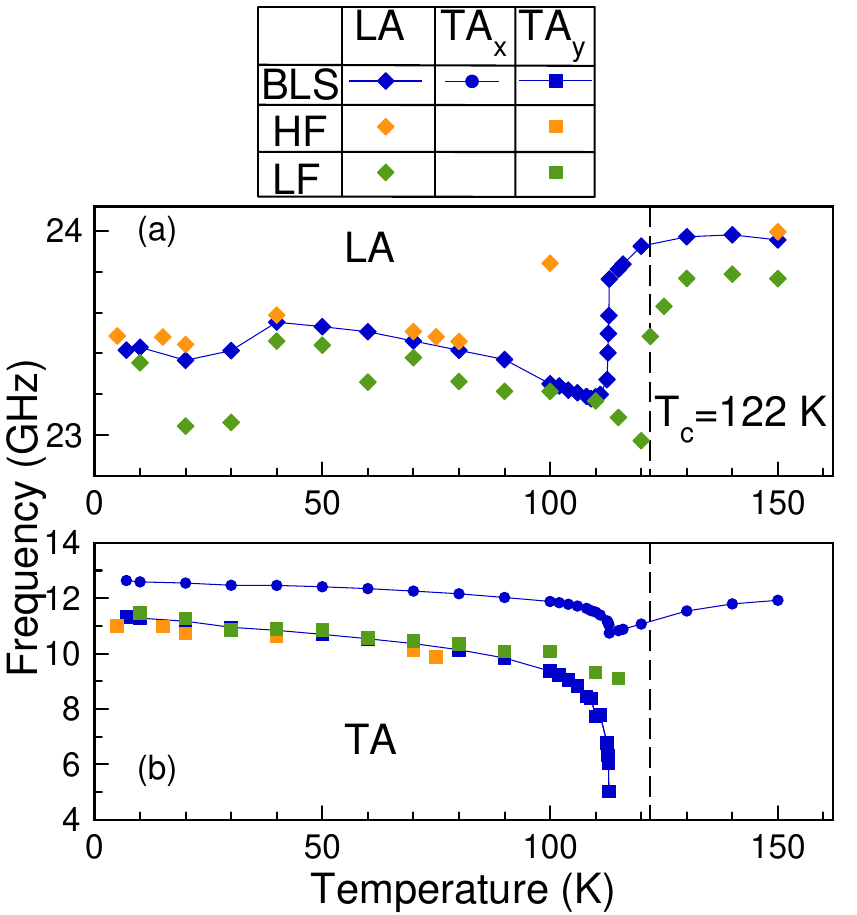}
\caption{{\bf Temperature dependence of the phonon frequencies.} The data are shown by orange symbols for the HF pump-probe, by the green symbols for the LF pump-probe, and by blue symbols for the $cw$ BLS spectroscopy (recalculated for the photon energy of $h\nu=2.34$ eV). 
(a) The diamond symbols correspond to the LA phonons, (b) squares to the TA$_{\rm y}$ phonons, circles to the TA$_{\rm x}$ phonons. The vertical dashed line corresponds to the phase transition temperature $T_c=122$~K.}
\label{fig:T-dep}
\end{figure}

The temperature dependence of the phonon frequencies and their amplitudes are particularly interesting because of the phase transition from tetragonal to cubic phase. The corresponding data are summarized in Fig.~\ref{fig:T-dep}. The phonon frequencies evaluated from the $cw$ BLS and LF pump-probe data are recalculated for the same probe photon energy of $h\nu=2.34$~eV used in the HF pump-probe measurements. Here, we assume a linear dependence of $f_i(h\nu)$ as follows from Fig.~\ref{fig:PumpProbeData}(d). This photon energy of the probe corresponds to maximum strength of the oscillatory signal, which happens close to the absorption edge, see Fig.~\ref{fig:generalInfo}(b). 

The $cw$ BLS data for the three acoustic phonon frequencies demonstrate sharp changes for temperatures around the structural phase transition at $T_c\sim 122$~K. For the longitudinal mode we observe an increase of frequency from 23.4~GHz by about 0.5 GHz with increasing temperature across the transition from the tetragonal to cubic phase. The frequency of the TA$_{\rm x}$ phonon undergoes a small decrease right below the phase transition, while above $T_c$ its magnitude recovers to a similar value as in the tetragonal phase. Interestingly, we observe a strong softening of the TA$_{\rm y}$ mode in the tetragonal phase when approaching the phase transition. The TA$_{\rm y}$ peak shifts from 11 to $5$~GHz accompanied by the formation of a broad band with practically zero offset (see Sec.~\ref{sec:si:BLS} of the Supplementary Information for $cw$ BLS spectra around the phase transition). In the cubic phase, the TA modes are degenerate so that only one peak corresponding to the TA phonons ($\sim 12$~GHz at $150$~K) is observed in the BLS spectrum.

The difference in the temperature at which the phase transition occurs in the $cw$ and time-domain BLS is attributed to local heating of the sample by the laser, which is most pronounced for HF excitation for which the integrated power is very high (2~kW/cm$^2$) so that the actual crystal temperature within the excitation spot is about 40~K higher than the bath temperature. For $cw$ BLS the excitation power is even higher (100~kW/cm$^2$) but the photon energy of 2.287~eV is below the direct band gap and the corresponding increase of temperature is only about 10~K. This is in contrast to the negligible heating observed in the LF pump-probe data with a time-integrated power of about $20$~W/cm$^2$, for which the phase transition temperature corresponds to the value of 122~K reported in literature~\cite{Schade2019}. 

As discussed in Sec.~\ref{sec:TDBLS}, only the LA and TA$_{\rm y}$ phonons are detected in the tetragonal phase  by time-domain BLS, while for $T>T_c$ in the cubic phase, only the LA coherent phonons are present in the spectrum. The temperature dependence of the phonon frequencies is in correspondence with those obtained in $cw$ BLS spectroscopy, i.e. we observe the hardening of the LA mode by 0.5~GHz and the softening of TA$_{\rm y}$ from 11 to 9~GHz in the vicinity of the phase transition (see Fig.~\ref{fig:T-dep}). Accordingly, the softening of the TA$_{\rm y}$ mode allows one to tune the speed of sound of the photoinduced shear strain pulse by varying the temperature. In addition, we note that the oscillation amplitudes for the LA and TA$_{\rm y}$ phonons remain practically the same in the whole temperature range from 5 to 150~K. The temperature dependence of the photoelastic constants in the transparency range is expected to be weak and, therefore, should not influence the strength of the detected signal. This implies that the magnitude of the photoinduced strain is independent of temperature and given by the density of photoexcited carriers.

\section*{Discussion}

To summarize, we have revealed and elaborated a new mechanism for efficient generation of shear strain in the tetragonal phase of lead-free double Cs$_2$AgBiBr$_6$ perovskite single crystals. The mechanism manifests in: (i) generation of coherent TA phonons with polarization dictated by the orientation of the $c$-axis; (ii) observation of soft TA phonons with a strong frequency dependence at temperatures around that of the structural phase transition; (iii) strong polarization dependence in the optical detection by a probe. The mechanism of shear strain generation is based on anisotropic photostriction of the crystal lattice so that it should be present in a large variety of perovskite materials with tetragonal crystal symmetry, e.g. in MAPbI$_3$ at room temperature. 

The weak temperature dependence of the photogenerated strain evidences the non-thermal nature of the effect, i.e., the direct generation of strain via carrier-phonon interaction accomplished due to the anisotropic photostriction. Indeed, the heat capacity has a strong nonlinear $T$-dependence, which possesses a sharp increase below 70~K and a slow variation at higher temperatures, while the thermal expansion coefficient does not depend on temperature~\cite{Schade2019}. 

Using the temperature dependence of the heat capacity from Ref.~\cite{Schade2019} and the absorption coefficient $\alpha=1.5\times 10^5$~cm$^{-1}$ from Ref.~\cite{Huygen2021}, we can estimate the increase of the temperature $\delta T$ after excitation with the maximum fluence of $\Psi_{\rm pump}=0.7$~mJ/cm$^2$. For instance, at $T=5$~K the optical excitation leads to $\delta T \approx 47$~K, while for $T=100$~K the temperature increase is significantly reduced to $\delta T(100\text{~K}) = 23$~K. Using the linear temperature expansion coefficient of $|\beta_i| \approx 10^{-5}$~K$^{-1}$, we estimate the amplitude of the strain field to be in the order of $10^{-3}$. Note that $\delta T(5\text{~K})/ \delta T(100\text{~K})\sim 2$ means that the temperature effect on the strain generation should result in a strong change (at least by a factor of two) of the signal amplitude when the temperature of the sample changes from 5 to 100~K. The weak dependence of the photoinduced strain on temperature in this range confirms that the main effect is not associated to the temperature-based mechanism (see Eq.~\eqref{eq:sigma_dT}) and occurs mainly due to the photostriction, given by Eq.~\eqref{eq:sigma_nX}. Assuming that the photostriction mechanism is at least as strong as that through the temperature change, we can estimate the photostriction coefficient $\alpha_i$. Using the exciton density $\rho_X(0) \approx 5\times 10^{19}$cm$^{-3}$ near the surface, we obtain the photostriction coefficient $|\alpha_i| \sim 10^{-23}$~cm$^3$.

We stress that the anisotropic photostriction is directly related to the strong interaction of electrons with optical phonons in combination with the crystallographic phase transition. Therefore, the excitation of soft TA coherent phonons could be due to interaction with the specific optical phonon eigenmodes responsible for the tetragonal to cubic phase transition. Particularly such optical phonons posses a strong interaction with charge carriers~\cite{Hoffman2023}. Further studies of the microscopic origin of the anisotropic photostriction could allow one to adjust the parameters of the material for efficient photoinduced generation of shear strain and realization of perovskite-based phononic transducers, including those for chiral phonons.

\section*{Methods}

{\bf Sample growth.} Single crystals of Cs$_{2}$AgBiBr$_{6}$ were grown by controlled cooling following the protocol in Refs.~\cite{Adam-2016, Armer-2021}. CsBr (1.5 mmol, 319.21 mg, 2 eq), AgBr (0.75 mmol, 140.83 mg, 1 eq), and BiBr$_{3}$ (0.75 mmol, 336.52 mg, 1 eq) were mixed in a 25 mL vial, followed by the addition of 10 ml of 48$\%$ wt HBr. The reaction was heated in a silicon oil bath to 110$^\circ$C during 20 min and kept at this temperature for 4 h to obtain a completely dissolved precursor solution. Next, the solution was cooled down at a rate of 1$^\circ$C/h to room temperature to obtain single crystals of approximately 4 mm size. The crystals were then removed from the mother liquor, dried with paper tissue, and rinsed with dichloromethane. Single crystals were grinded into powder prior to X-ray diffraction (XRD) measurements. 

{\bf Sample characterization.} XRD measurements were carried out using a General Electric XRD 3003 TT diffraction system using a Cu-K$_{\alpha}$ radiation source with a wavelength $\lambda$ of 1.5406 \r{A} $(V = 40$ kV, $I = 40$ mA), in ambient conditions. The resulting XRD pattern of Cs$_{2}$AgBiBr$_{6}$ matches the simulated data for a cubic lattice \cite{doi:10.1021/jacs.7b01629} (see Supplementary Information, Fig.~\ref{fig:XRD}).

{\bf Photoluminescence spectroscopy.} The PL spectra were measured at $T=2$~K in a bath cryostat with the sample immersed in superfluid helium. The photon energy for excitation is $3.493$ eV. A halogen lamp was used as a source of white light for reflection measurements. 

{\bf Time-domain and $cw$ Brillouin spectroscopy.} In pump-probe and $cw$ BLS studies the sample was cooled down in a helium flow cryostat, allowing the use of microscope objectives with working distance $>10$~mm. In all experiments a $20\times$ microscope objective with a numerical aperture of 0.4 was used. The beams were focused into spots of about 5~$\mu$m in diameter. 

Continuous wave ($cw$) Brillouin light scattering was measured using a stabilized double tandem Fabry-P\'erot Brillouin spectrometer (TFP-2 from Table Stable). The distance between the mirrors is set to 3 mm, and the width of the entrance and exit slits was $300$~$\mu$m. For excitation we used a single frequency $cw$ laser with the photon energy of $2.287$ eV ($542$ nm wavelength) and a spectral width below 10~MHz.  The sample was excited with the laser power of 20 mW in backscattering geometry with nominally normal incidence (unless specified otherwise). The scattered signal was detected in a co-polarized scheme, corresponding to the detection of vertically polarized backscattered light.

Time-domain Brillouin spectroscopy was performed using a transient pump-probe technique at two different repetition frequencies of 80~MHz and 30~kHz. The pump-probe at 80~MHz was implemented using an asynchronious optical sampling technique with two laser sources tunable in the range of $470\div 700$~nm wavelength (Toptica  FFpro). The repetition rates of the lasers were synchronized with an offset frequency equal to $\delta f = 2$ kHz. The sample was excited by both lasers serving as pump and probe, using the same microscope objective in confocal reflection geometry. The reflected probe beam was detected using a fast photodiode in combination with a digitizer card, giving an overall time resolution better than $10$~ps. The energy fluence of the pump and probe pulses $\Psi$ were set to 30~$\mu$J/cm$^2$ and 10~$\mu$J/cm$^2$, respectively. 

Time-domain Brillouin spectroscopy at 30~kHz was performed using an optical parametric amplifier system tunable in the range $0.4\div 2$~$\mu$m (PHAROS Femtosecond Lasers). A delay line was introduced using a retroreflector mounted on a  motorized linear stage. The intensity of the beam was modulated with a chopper at the frequency of $1$~kHz and the reflected probe beam was detected with a photodiode in conjunction with a lock-in amplifier. The energy fluences of the pump and probe pulses $\Psi$ were set to 0.7~mJ/cm$^2$ and 3.5~mJ/cm$^2$, respectively. Even though the fluence of the probe pulse exceeded that of the pump pulse, its photon energy remained below the energy gap ($h\nu_{\rm probe}<2$~eV). Consequently, the probe beam caused negligible absorption that could influence the density of photo-excited carriers. 

The duration of the pulses in both pump-probe setups was about $150$~fs. The angular dependence of polarization in Fig.~\ref{fig:PumpProbeData}(b) was taken for linearly co-polarized pump and probe beams. The polarization of the beams was set using Glan-polarizers. The polarization rotation was achieved using the half-wave plate in front of the microscope objective which corresponds to a rotation of the sample with respect to the laboratory frame.

{\bf Density functional theory calculations.} The DFT calculations were performed using the WIEN2k package \cite{Blaha2020}. For the calculation of the elastic tensor components we use the IRelast package \cite{Jamal2018}. First, the structure is optimized using the PBEsol exchange-correlation functional. Next, for the few selected deformations the energy as a function of deformation amplitude is calculated and from the fit of the calculated energy, combinations of elastic tensor components are extracted.

{\bf Data availability.} The data on which the plots within this paper are based and other findings of this study are available from the corresponding author upon justified request.

\subsection*{ORCID} 
\noindent Dmytro Horiachyi: 0009-0002-1510-5414\\
Mikhail O. Nestoklon: 0000-0002-0454-342X\\
Ilya A. Akimov:   0000-0002-2035-2324    \\
Artur V. Trifonov : 0000-0002-3830-6035\\
Nikita V. Siverin: 0000-0002-4643-845X\\
Nataliia E. Kopteva:   0000-0003-0865-0393 \\
Alexander N. Kosarev: 0000-0002-5201-6923\\
Dmitri R. Yakovlev:   0000-0001-7349-2745 \\
Vitalyi E. Gusev: 0000-0002-2394-7892\\
Melina Fries: 0000-0002-1716-4648\\
Olga Trukhina: 0009-0000-9709-4872\\
Vladimir Dyakonov: 0000-0001-8725-9573\\
Manfred Bayer:        0000-0002-0893-5949\\

\bibliographystyle{unsrt}
\bibliography{refs}

\onecolumngrid

\subsection{Acknowledgments}
The authors are grateful to S.V.~Goupalov, A.V.~Akimov, M.M.~Glazov, and A.V.~Scherbakov for useful discussions. The Dortmund and W\"urzburg groups acknowledge financial support from the Deutsche Forschungsgemeinschaft within the framework of the SPP 2196 (Project No. 506623857). D.O.H. acknowledges the Deutsche Forschungsgemeinschaft (Project No. $536987509$).
M.O.N. gratefully acknowledges the computing time provided on the Linux HPC cluster at Technical University Dortmund (LiDO3), partially funded in the course of the Large-Scale Equipment Initiative by the Deutsche Forschungsgemeinschaft (Project No. 271512359). N.V.S. and D.R.Y. acknowledge financial support of the Deutsche Forschungsgemeinschaft via the Collaborative Research Center TRR142 (project A11).

\subsection{Author contributions}
D.O.H., A.V.T., N.V.S., and A.N.K. built the experimental apparatus and performed the measurements.
I.A.A and D.H. analyzed the data.
N.E.K. measured the PL, transmission, and absorption spectra.
M.O.N. and V.E.G developed the theoretical approach and performed the model calculations.
M.F., O.T. and V.D. grew the samples and performed XRD. 
All authors contributed to interpretation and analysis of the data. 
D.O.H, I.A.A and M.O.N. wrote the manuscript in close consultations with M.B., V.E.G, D.R.Y. and V.D.

\subsection{Additional information}
Correspondence should be addressed to D.O.H. (email: dmytro.horiachyi@tu-dortmund.de ) and M.O.N. (email: mikhail.nestoklon@tu-dortmund.de).

\subsection{Competing financial interests}
The authors declare no competing financial interests.

\clearpage
\newpage
\begin{center}
  \textbf{{\Large Supplementary Information:}}\\
  \textbf{{\Large Efficient launch of shear phonons in photostrictive halide perovskites}}
  \\
 \vspace {7mm}
D.O.~Horiachyi, M.O.~Nestoklon, I.A.~Akimov, A.V.~Trifonov, N.V.~Siverin, N.E.~Kopteva, A.N.~Kosarev, D.R.~Yakovlev, V.E.~Gusev, M.~Fries, O.~Trukhina, V.~Dyakonov, and M.~Bayer
\end{center}
\hypersetup{pageanchor=false}  

\setcounter{equation}{0}
\setcounter{figure}{0}
\setcounter{table}{0}
\setcounter{page}{1}
\setcounter{section}{0}
\renewcommand{\thepage}{S\arabic{page}}
\renewcommand{\theequation}{S\arabic{equation}}
\renewcommand{\thefigure}{S\arabic{figure}}
\renewcommand{\thetable}{S\arabic{table}}
\renewcommand{\thesection}{S\arabic{section}}
\renewcommand{\theHsection}{Ssection.\arabic{section}}
\renewcommand{\bibnumfmt}[1]{[S#1]}
\renewcommand{\citenumfont}[1]{S#1}
\renewcommand{\theHequation}{Sequation.\arabic{equation}}  
\renewcommand\theHfigure{Sfigure.\arabic{figure}}  

\section{X-ray diffraction}\label{sec:si:XRD}

\begin{figure}[ht]
\includegraphics[width=0.5\linewidth]{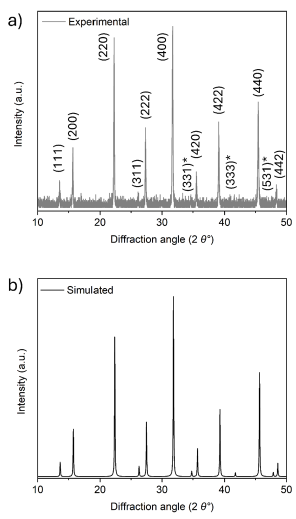}
\caption{(a) XRD powder pattern of the grained Cs$_{2}$AgBiBr$_{6}$ crystals measured at room temperature. The peaks can be assigned to a simple cubic crystal structure~\cite{Adam-2016}. Peaks expected, but not resolved are marked with asterisks. (b) Simulated XRD powder pattern of the Cs$_{2}$AgBiBr$_{6}$ perovskite. The simulated data were generated with the Mercury CCDC software using the crystallographic data from~\cite{doi:10.1021/jacs.7b01629}.} 
\label{fig:XRD}
\end{figure}

The results of X-ray diffraction measurements and corresponding simulations for grained Cs$_{2}$AgBiBr$_{6}$ crystals are shown in Fig.~\ref{fig:XRD}.

\section{Continuous wave Brillouin light scattering spectroscopy}
\label{sec:si:BLS}

\begin{figure}[ht]
\centering
\includegraphics[width=0.7\linewidth]{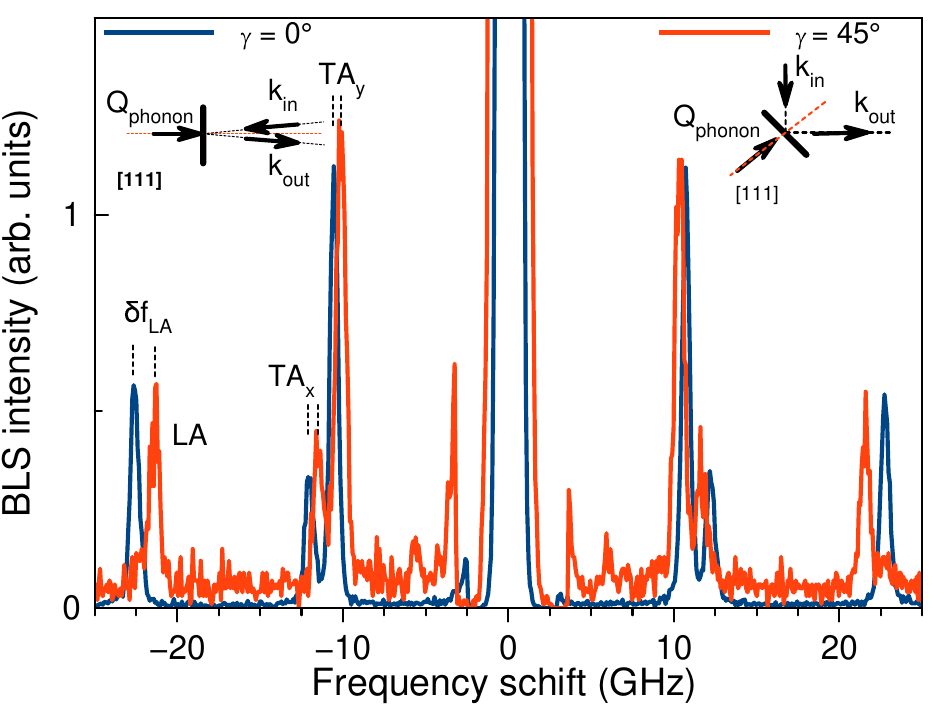}
\caption{Spectra of Brillouin light scattering (BLS) in back reflection geometry for incidence angles of $\gamma=0^\circ$ (normal incidence, blue) and $\gamma=45^\circ$ (red). The insets show schematic presentations of the experimental geometries. 
$\mathbf{k_{in}}$ and $\mathbf{k_{out}}$ are the wavevectors of the incident and scattered light, respectively, $\mathbf{Q}$ is the wavevector of the phonon involved in the scattering process, ${[111]}$ is the crystallographic orientation axis of the sample. $T=5$~K.}
\label{fig:BLS}
\end{figure}

The positions of the BLS peaks measured at $T=5$~K in reflection geometry for two geometries, normal light incidence and $45^\circ$ light incidence with respect to the $x'y'$ sample surface, are compared. The results are presented in Fig.~\ref{fig:BLS}. The change of the incidence angle in reflection geometry leads to a variation of the wavevector of the phonon involved in scattering $q \approx 2k_{in}\sin{(\Theta/2)}$ with $k_{\rm in} \approx k_{\rm out}$, where $k_{\rm in}$ and $k_{\rm out}$ are the wavevectors of the incident and scattered light and the angle $\Theta = \pi-\arcsin{(\sin{(\gamma)}/n_r)}$ is determined by the incidence angle $\gamma$. Note that in this case the direction of the wavevector $\mathbf{Q}$ remains orthogonal to the crystal surface, i.e. the sound velocities are equal for both geometries. Taking into account the refractive index of the Cs$_{2}$AgBiBr$_{6}$ perovskite, we can calculate the relative frequency shift of the lines for two geometries: $\delta f_i(\gamma) = [f_i(\gamma)- f_i(0)]/f_i(0)$ which corresponds to 5.5\% for $\gamma=45^\circ$. Experimentally we observe shifts of the frequencies equal to 5.3\%, 4.4\%, and  3.8\% for the longitudinal and two transverse acoustical phonons, respectively.

The polarization dependence of the BLS peaks with respect to the crystallographic axes allows us to determine the direction of the c-axis. We observe that the c-axis is randomly oriented along one of three equivalent directions after each cooling cycle from the cubic to the tetragonal crystallographic phase. However, once cooled below $T_c$, the relative intensities of the BLS peaks remain the same when examining different areas of the sample. This suggests that the direction of the c-axis does not change from one point to another, enabling us to rule out the presence of multiple domains with varying c-axis orientations in the tetragonal phase.

\begin{figure}[ht]
\includegraphics[width=0.8\linewidth]{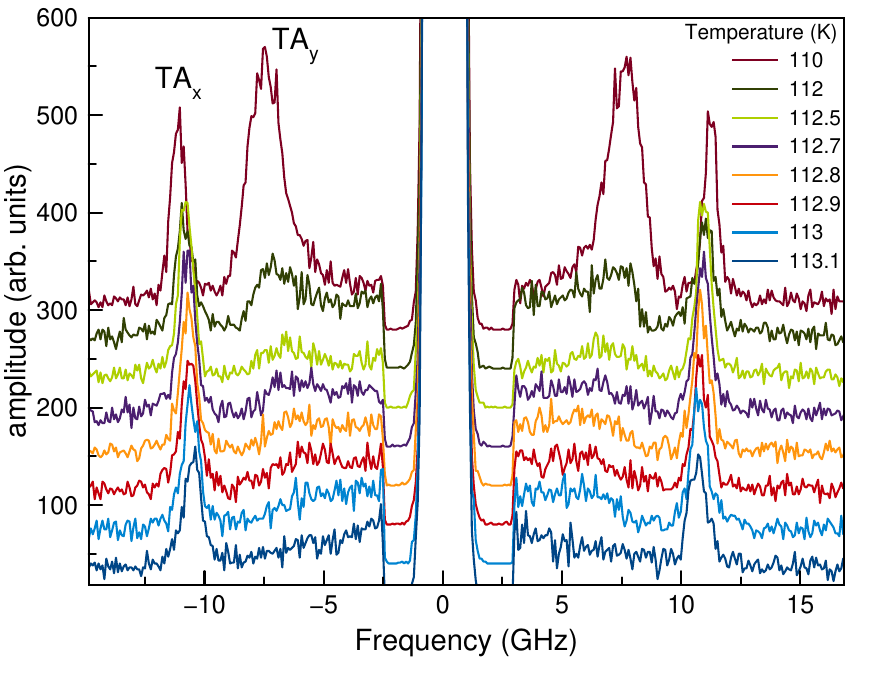}
\caption{Temperature dependence of the $cw$ BLS spectra close to the phase transition.} 
\label{fig:Softening}
\end{figure}

Close to the phase transition temperature, softening of the transverse acoustic modes is observed. With increasing temperature the the acoustic mode frequencies decrease. The softening is particularly strong for the TA$_y$ mode where the BLS shift tends to zero frequency, accompanied by a strong broadening and vanishing of the signal with increasing temperature (see Fig.~\ref{fig:Softening}). 

\section{Material tensors}\label{sec:si:tensors}

The tensor of elastic stiffness is defined \cite{ITCBD} as 
\begin{equation}\label{eq:c_def}
  \sigma_{ij} = c_{ijkl} \epsilon_{kl}\,,
\end{equation}
where $\sigma_{ij}$ is the stress tensor and $\epsilon_{kl}$ is the strain tensor. We use the standard convention where the symmetric stress and strain tensors are written as six-component vectors associating pairs of tensor indices $ij$ with a single index $\zeta$ like $11\to1$, $22\to2$, $33\to3$, $23\to4$, $13\to5$, $12\to6$. With this convention, the rank four symmetric tensor (tensor of elastic stiffness and photoelastic tensor) can be written as a $6\times6$ matrix. 

The photoelastic (opto-elastic) tensor is defined \cite{ITCBD} through the linear dependence of the components of the dielectric impermeabilities (inversed dielectric tensor) on the components of strain tensor:
\begin{equation}\label{eq:p_def}
  \delta \left\{\varepsilon^{-1}\right\}_{ij} = p_{ijkl} \epsilon_{kl}\,,
\end{equation}
where $p_{ijkl}$ are the components of the photoelastic tensor. 
For an isotropic dielectric constant $\varepsilon_{ij} = \varepsilon \delta_{ij}$, the change of the dielectric constant is proportional to the same tensor:
\begin{equation}\label{eq:dn}
  \delta \varepsilon_{ij} = \varepsilon^2 p_{ijkl} \epsilon_{kl}\,.
\end{equation}

In the cubic phase, the space group of the crystal is \#225 and the elastic stiffness tensor and photoelastic tensor have the same form:
 \begin{equation}\label{eq:cp_225}
   C_{ij} = \begin{pmatrix}
    c_{11} & c_{12} & c_{12} & 0 & 0 & 0 \\
    c_{12} & c_{11} & c_{12} & 0 & 0 & 0 \\
    c_{12} & c_{12} & c_{11} & 0 & 0 & 0 \\
    0 & 0 & 0 & c_{44} & 0 & 0 \\
    0 & 0 & 0 & 0 & c_{44} & 0 \\
    0 & 0 & 0 & 0 & 0 & c_{44} \\
   \end{pmatrix}\,,\;\;\;
  p_{ij} = \begin{pmatrix}
    p_{11} & p_{12} & p_{12} & 0 & 0 & 0 \\
    p_{12} & p_{11} & p_{12} & 0 & 0 & 0 \\
    p_{12} & p_{12} & p_{11} & 0 & 0 & 0 \\
    0 & 0 & 0 & p_{44} & 0 & 0 \\
    0 & 0 & 0 & 0 & p_{44} & 0 \\
    0 & 0 & 0 & 0 & 0 & p_{44} \\
   \end{pmatrix}\,.
 \end{equation}
Note that in contrast to the main text, in this section and below we use the standard convention and write these four-rank tensors as matrices $6\times 6$.

The tetragonal phase has the space group \#87, class $4/m$, and the tensors are different. Both may be found in \cite{ITCBD}, \S~1.1.4.10.6.5 (Note that there is an error in Table~1.6.7.1 of the same volume) 
\begin{equation}\label{eq:cp_87}
  C_{ij} = \begin{pmatrix}
   c_{11} & c_{12} & c_{13} & 0 & 0 & c_{16} \\
   c_{12} & c_{11} & c_{13} & 0 & 0 &-c_{16} \\
   c_{13} & c_{13} & c_{33} & 0 & 0 & 0 \\
   0 & 0 & 0 & c_{44} & 0 & 0 \\
   0 & 0 & 0 & 0 & c_{44} & 0 \\
   c_{16} & -c_{16} & 0 & 0 & 0 & c_{66} \\
  \end{pmatrix}\,,\;\;\;
 p_{ij} = \begin{pmatrix}
   p_{11} & p_{12} & p_{13} & 0 & 0 & p_{16} \\
   p_{12} & p_{11} & p_{13} & 0 & 0 &-p_{16} \\
   p_{31} & p_{31} & p_{33} & 0 & 0 & 0 \\
   0 & 0 & 0 & p_{44} & p_{45} & 0 \\
   0 & 0 & 0 &-p_{45} & p_{44} & 0 \\
   p_{61} & -p_{61} & 0 & 0 & 0 & p_{66} \\
  \end{pmatrix}\,.
\end{equation}
It is worth to mention that, as explained in Refs.~\cite{Nelson70,Anastassakis74}, a change of the dielectric constant in low symmetry crystals is not fully captured by the tensor $p_{ij}$, but contains also an antisymmetric part which describes the change of the refraction index by twist stress. However, as explained in \cite{Vacher72}, these terms are zero if the refractive index is isotropic. According to the experimental data, the anisotropy of the refractive index in our case may be neglected and the photoelastic interaction may be described by the tensor \eqref{eq:cp_87}.

To relate the cubic phase components to the tetragonal ones, an additional step is needed. With the standard crystallographic basis choice, the (normalized) basis vectors in the tetragonal phase $\mathbf{e}^{\rm t}_{j}$ are related to the basis vectors in the cubic phase $\mathbf{e}^{c}_{j}$ by $\mathbf{e}^{\rm t}_{1} = (\mathbf{e}^{c}_{1}-\mathbf{e}^{c}_{2})/\sqrt2$, $\mathbf{e}^{\rm t}_{2} = (\mathbf{e}^{c}_{1}+\mathbf{e}^{c}_{2})/\sqrt2$, $\mathbf{e}^{\rm t}_{3} = \mathbf{e}^{c}_{3}$. From comparison of the rotated cubic tensor of elastic stiffness with the tetragonal one, one may conclude that the deviation of the tetragonal stiffness from the cubic material with parameters $c_{ij}^c$ (where $c_{11}^c = c_{33}$, $c_{12}^c = c_{13}$, and $c_{44}^c =c_{44}$) may be expressed via 
\begin{equation}
  \delta c_{66} = c_{66}-\frac{c_{33}-c_{13}}2\,,\;\;\;
  \delta c_{11} = c_{11}-\left(c_{44}+\frac{c_{33}+c_{13}}2\right)\,,\;\;\;
  \delta c_{12} = c_{12}+\left(c_{44}-\frac{c_{33}+c_{13}}2\right)\,,\;\;\;
  c_{16}\,.
\end{equation}

Note also that the $\beta$-phase of perovskites has a slightly higher symmetry (space group \#140, class $4/mmm$), where $p_{16}=p_{61}=p_{45}=0$ (we remind that there is an error in Ref.~\citenum{ITCBD} in Table 1.6.7.1.).

Since the elastic stiffness tensor is not known for this material, we performed density functional theory (DFT) calculations to estimate the mode mixing in bulk material. The DFT calculations were done using the WIEN2k package \cite{Blaha2020}. For calculation of the elastic tensor components we use the IRelast package \cite{Jamal2018}. First, the structure is optimized using the PBEsol exchange-correlation functional (the optimized cubic lattice constant is 11.19~\AA\ and in the tetragonal phase $a=7.84$~\AA, $c=11.32$~\AA). Next, for a few selected deformations, the energy as function of deformation amplitude is calculated and from the fit of the calculated energy the combinations of the elastic tensor components are extracted. The procedure gives the following set of constants in the cubic phase (constants are in GPa):
\begin{equation}
  c_{11} = 45.7\,,\,\, 
  c_{12} = 17.2\,,\,\,
  c_{44} =  7.0\,,
\end{equation}
and in the tetragonal phase:
\begin{equation}\label{eq:CtDFT}
\begin{split}
  c_{11} = 41.0,\,\, 
  c_{33} = 42.7,\,\, \\
  c_{44} =  7.10,\,\, 
  c_{66} = 12.5,\,\, \\
  c_{12} = 26.2,\,\,
  c_{13} = 22.5,\,\,
  c_{16} = -0.5. 
\end{split}
\end{equation}

\section{Phonon modes for tetragonal phase}\label{sec:si:phonons}

For qualitative analysis it is practical to write the phonon dispersion equation in the tetragonal phase in the coordinate system aligned with the $\mathbf{z}'$ direction. 
In the coordinates $x'y'z'$ it reads as
\begin{equation}\label{eq:disp_ph_111}
  \left\{ C_{ijkl} Q_j Q_k \right\}_{x'y'z'} = \frac13 
  \begin{pmatrix}
  c_{T}+2\delta c_{66}      & -\frac{2}{\sqrt3} c_{16} & -\frac{4}{\sqrt6} c_{16} \\
  -\frac{2}{\sqrt3} c_{16} & c_{T}+\frac23 \delta c_{11} & \frac{2\sqrt2}3 \delta c_{11} \\
  -\frac{4}{\sqrt6} c_{16}  &  \frac{2\sqrt2}3 \delta c_{11} & c_L + \frac43 \delta c_{11}
 \end{pmatrix}\,.
\end{equation}
where
\begin{align*}
  c_{T}  &= c_{33}-c_{13}+c_{44}\,,\,\,\,\\
  c_{L}  &= c_{33}+2c_{13} + 4c_{44} \,,\,\,\, \\
\end{align*}
In the cubic limit, the dispersion equation gives two degenerate transverse phonon modes $V_{\mathrm TA1}=V_{\mathrm TA2}\propto c_{33}-c_{13}+c_{44}$ and one longitudinal mode $\propto c_{33}+2c_{13} + 4c_{44}$, as expected for cubic material. 
Below we fix the notation as follows: The phonon mode TA$_1$ is the mode polarized along $\mathbf{x}'$ (normal to the $c$ axis) in the cubic limit, TA$_2$ is the mode polarized along $\mathbf{y}'$ (non-zero projection on the $c$ axis).

There are three effects of the tetragonal anisotropy on the polarization vectors ${\bf U}^s$: (1) there is an admixture of the LA mode to the TA$_2$ mode, that is proportional to $\delta c_{11}$, (2) there is an admixture of the LA mode to the TA$_1$ mode that is proportional to $c_{16}$, and (3) there is a mixing of the TA$_1$ and TA$_2$ modes, also proportional to $c_{16}$. Note that each of these admixtures is reflected both in the velocities and in the displacement vectors ${\bf U}^s$. An admixture of LA phonons means that the TA modes are no longer purely transverse, but have a component of the displacement vector along ${\bf Q}$. Mixing of the transverse phonons means that the main axes of transverse phonons is rotated with respect to $\mathbf{x}'$ and $\mathbf{y}'$ in the plane $x'y'$. Exact calculations with the stiffness tensor components extracted from DFT allows us to estimate the mixing of the different modes and shows that the effect of elastic anisotropy may be calculated perturbatively. 

An approximate solution of Eq. \eqref{eq:disp_ph_111} may be written explicitly taking into account that $\delta c_{11}\ll c_{L}-c_T$ and $c_{16}\ll \delta c_{66} - \delta c_{11}$. For our material this assumption works pretty well. 
\begin{equation}\label{eq:Us_approx}
  {\bf U}^{T1} \approx C_n^{T1} \begin{pmatrix} 1 \\ \sqrt3 \frac{c_{16}}{\delta c_{66}-\delta c_{11}} \\ 4\sqrt3 \frac{c_{16}}{c_{T}-c_{L}} \end{pmatrix}\,,\,\,\,
  {\bf U}^{T2} \approx C_n^{T2} \begin{pmatrix} \sqrt3 \frac{c_{16}}{\delta c_{66}-\delta c_{11}} \\ 1 \\ \frac{2\sqrt2}3 \frac{ \delta c_{11}}{c_{L}-c_{T}} \end{pmatrix}\,,\,\,\,
  {\bf U}^{L } \approx C_n^{L}  \begin{pmatrix} \frac{4}{\sqrt3} \frac{c_{16}}{c_{L}-c_{T}} \\ \frac{2\sqrt2}3\frac{\delta c_{11}}{c_{L}-c_{T}} \\1 \end{pmatrix}\,,\,\,\,
\end{equation}
where the $C_n^s$ are the normalization coefficients.

From Eq. \eqref{eq:Us_approx} one may qualitatively understand the behavior of the energies and displacement vectors. The ``zeroth order'' modes contain squares of velocities proportional to $c_{T}+2/3\,\delta c_{66}$, $c_{T}+ 2/3\,\delta c_{11}$ and $c_L$ with their displacement vectors aligned: TA$_1$ along $\mathbf{x}'$, TA$_2$ along $\mathbf{y}'$, and LA along $\mathbf{z}'$. Then, in the first order, the displacement vectors are rotated in the $x'z'$ plane proportional to $c_{16}$, in the $x'y'$ plane also proportional to $c_{16}$, and in the $y'z'$ plane proportional to $\delta c_{11}$.

The phonon amplitudes ${\bf U}^s$ may be recalculated to strain tensor components which enter both the Brillouin scattering and pump-probe Brillouin amplitude:
\begin{equation}\label{eq:eps_0}
  \epsilon^s_{ij} = \frac12 (U_i^s Q_j + U_i^s Q_i)\,.
\end{equation}
In the coordinate system $x'y'z'$, only $Q_{z'}$ is non-zero. 
As a result, only three components of the strain tensor are non-zero, namely
\begin{equation}\label{eq:str_amp}
 \epsilon^s_{z'z'} = U_{z'}^s\,,\,\,\,
 \epsilon^s_{y'z'} = \frac12 U_{y'}^s\,,\,\,\,
 \epsilon^s_{z'x'} = \frac12 U_{x'}^s\,.
\end{equation}
In our case, the modes of interest are the $LA$ phonons with the dominatig component of the strain tensor $\epsilon_{z'z'}$ and the $TA_{2}$ phonons with the dominating component $ \epsilon_{y'z'}$

\section{Generation of strain wave}\label{sec:si:strain_wave}
Now let us obtain a more formal solution of the problem. The wave equation with the right part reads
\begin{subequations}\label{eq:we}
\begin{align}
  \rho \frac{d^2 u_i}{dt^2} &= \dd{\sigma_{ij}}{x_j}\,,\label{eq:dudt} \\
  \sigma_{ij} &= c_{ijkl} \epsilon_{kl} + \Sigma_{ij}\,,\label{eq:sigma}\\
  \epsilon_{kl} &= \frac12 \left( \dd{u_k}{x_l} + \dd{u_l}{x_k} \right)\,. \label{eq:epsilon}
\end{align}
\end{subequations}
Here $\Sigma_{kl}$ is the stress induced by the laser light, $\epsilon$ is the deformation tensor, and ${\bf u}$ is the displacement vector.

Equation \eqref{eq:eigen_ph} is the Fourier transform of \eqref{eq:we} without the right part. The Fourier transform of the original problem with the right part included gives 
\begin{equation}\label{eq:eigen_ph_F}
  \left( \rho\omega^2 \delta_{il} - c_{ijkl} k_j k_k \right) U_l({\bf k},\omega) = \Phi_i({\bf k},\omega)\,,
\end{equation}
with the right part Fourier transform of the driving force $\mathbf{\Phi}$: 
\[
  \Phi_i({\bf k},\omega) = \int \Phi_i({\bf r},t) e^{-i({\bf k}\cdot{\bf r} - \omega t)} d{\bf r} dt.
\]
Without loss of generality, below we give results only for the photostriction contribution \eqref{eq:sigma_nX}.
Direct calculations show that 
\begin{equation}
  \Phi_i({\bf k},\omega) =  -n_i \alpha_i \rho_{X,S} \left[ 2\pi \delta(\omega) - \frac{2i}{\omega} \right] \frac{2 n_x n_y n_z\zeta}{(n_x^2+k_x^2)(n_y^2+k_y^2)(n_z^2+k_z^2) }.
\end{equation}
It is clear that the solution of \eqref{eq:eigen_ph_F} gives three waves propagating with velocities corresponding to acoustic eigenmodes, and the amplitude of the mode $s$ is proportional to ${\bm{\Phi}}\cdot {\bf U^s}$.

The displacement field in the wave after the pulse is proportional to 
\begin{equation}
  u_i({\bf r},t) \sim \sum_{s={\mathrm LA}, {\mathrm TA_x}, {\mathrm TA_y}} A_s U_i^s f\left( t-\frac{ {\bf r}\cdot{\bf n}}{V_s} \right) \,,
\end{equation}
where $f$ is the envelope of the strain pulse. The strain tensor is 
\begin{equation}\label{eq:str_tensor_ampl}
  \epsilon_{ij} ({\bf r},t) = -\sum_{s={\mathrm LA}, {\mathrm TA_x}, {\mathrm TA_y}}\frac{A_s}{2V_s} f'\left( t-\frac{ {\bf r}\cdot{\bf n}}{V_s} \right) \left( U_i^s n_j + U_j^s n_i \right).
\end{equation}
Note that the strain tensor for different modes is proportional not only to the amplitude $A_s \sim {\bm{\Phi}}\cdot {\bf U}^s$ but also to $\epsilon^s_{ij} = \frac12 (U_i^s Q_j + U_i^s Q_i)$. The polarization dependence of the signal is defined only by $\epsilon^s_{ij}$ and the components of the photoelastic tensor, see the details in \ref{sec:si:phonons}.

\section{Detection}\label{sec:si:detect}
It can be demonstrated that the electric field of the scattered light can be written (see e.g. Refs.~\cite{Subbaswamy1978,Matsuda2008}) in first order as
\begin{equation}
  {\bf E}(z) \approx {\bf E}_0(z) + k^2 \int_{-\infty}^{\infty} G(z,z') \delta \varepsilon {\bf E}_0(z) {\rm d}z',
\end{equation}
where $\delta \varepsilon$ is the variation of the dielectric constant proportional to the strain \eqref{eq:dn} and $ G(z,z')$ is the Green function of the wave equation 
\begin{equation}
  \left[\dd{}{z} + k^2\varepsilon \right] G(z,z') = \delta(z-z')\,.
\end{equation}
In the 1D case with a constant dielectric function, it is given by
\begin{equation}
  G(z,z') = -\frac{i}{2k} e^{ik|z-z'|} \,,
\end{equation}
and is independent of the polarization of light. As a result, the amplitude of the scattered light is simply proportional to the amplitude of the dielectric tensor modulation $\delta \varepsilon_{ij}$, $i,j = x',y'$. 

The photoelastic tensor should be rewritten in terms of the coordinates $x'y'z'$. Then, the change of components of the dielectric tensor is proportional to its 3rd column for the LA phonon ($\epsilon_{z'z'}$), to the 4th column for the TA$_{2}$ phonon ($\epsilon_{y'z'}$), and to the 5th column for the TA$_{1}$ ($\epsilon_{x'z'}$). Since the light propagates along the $\mathbf{z}'$ direction, for the calculation of the reflected light we need only the following components of the dielectric tensor: $\delta \varepsilon_{x'x'}$, $\delta \varepsilon_{y'y'}$, $\delta \varepsilon_{x'y'}$. The result for a tetragonal $4/m$ crystal is the following:
\begin{subequations}\label{eq:polaLA}
\begin{align}
  \delta \varepsilon^{LA}_{x'x'} \propto \, & 2p_{12}+p_{13}\,,\\
  \delta \varepsilon^{LA}_{y'y'} \propto \, & (2p_{11}+2p_{33}+p_{13}+4p_{31}-8 p_{44})/3\,, \\
  \delta \varepsilon^{LA}_{x'y'} \propto \, & 2( p_{45} - p_{61})/\sqrt{3}\,,
\end{align}
\end{subequations}
\begin{subequations}\label{eq:polaTAy}
\begin{align}
  \delta \varepsilon^{TA_{2}}_{x'x'} \propto \, & p_{12}-p_{13}\,,\\
  \delta \varepsilon^{TA_{2}}_{y'y'} \propto \, &  (p_{11}-2p_{33}-p_{13} +2p_{31}+2p_{44})/3\,,\\
  \delta \varepsilon^{TA_{2}}_{x'y'} \propto \, & -( p_{45} + p_{16})/\sqrt3\,,
\end{align}
\end{subequations}
\begin{subequations}\label{eq:polaTAx}
\begin{align}
  \delta \varepsilon^{TA_{1}}_{x'x'} \propto \, & p_{16}\,,\\
  \delta \varepsilon^{TA_{1}}_{y'y'} \propto \, & -(2p_{45}+p_{61})/3\,,\\ 
  \delta \varepsilon^{TA_{1}}_{x'y'} \propto \, & (p_{66}-p_{44})/\sqrt3\,.
\end{align}
\end{subequations}
Let us discuss qualitatively the result o Eqs.~(\ref{eq:polaLA}-\ref{eq:polaTAx}). We consider the sequence 1. isotropic case $\rightarrow$ 2. cubic $m\bar3m$ $\rightarrow$ 3. tetragonal $4/mmm$ $\rightarrow$ 4. tetragonal $4/m$. 

1. In the fully isotropic case there are two independent components $p_{11}$ and $p_{12}$, while all others $p_{11}=p_{33}$, $p_{12}=p_{13}=p_{31}$, $p_{44}=p_{66}=\frac{p_{11}-p_{12}}{2}$, $p_{45}=p_{16}=p_{61}=0$ which leads to a polarization-independent reflection only for the LA phonons, while no signal from TA phonons can be detected even if they were generated.

2. In the cubic symmetry, there is an additional non-zero component $p_c^c = p_{11}^c-p_{12}^c-2p_{44}^c$ and Eqs.~(\ref{eq:polaLA}-\ref{eq:polaTAx}) reduce to:
\begin{subequations}\label{eq:polacub}
\begin{align}
  \delta \varepsilon^{LA}_{x'x'} = \delta \varepsilon^{LA}_{y'y'} \propto \, & 3p_{12}+p_c^c\,,\\
  \delta \varepsilon^{TA_{2}}_{x'x'} = -\delta \varepsilon^{TA_{2}}_{y'y'} \propto \, & p_c^c\,,\\
  \delta \varepsilon^{TA_{1}}_{x'y'} = \delta \varepsilon^{TA_{1}}_{y'x'}  \propto \,& p_c^c \,.
\end{align}
\end{subequations}
Eqs.~\eqref{eq:polacub} show that the signal for the LA phonons is still isotropic, and the TA phonons give a polarization dependent four-lobe signal with the maximum/minimum along the $x'$, $y'$ axes for the TA$_{2}$/TA$_{1}$ phonons (i.e., the signal is proportional to $|\cos 2 \phi|$, $|\sin 2\phi|$, respectively). Note that the amplitude of reflection for both TA phonons is given by the same constant $p_c$.

3. In the $4/mmm$ system, there are 7 independent photoelastic tensor components, only $p_{45}=p_{16}=p_{61}=0$. In this case, the TA$_{1}$ signal is expected to be the same as in the cubic case. To explicitly show the difference from the cubic symmetry, we write the result for the LA and TA$_{2}$ phonons as: 

\begin{subequations}\label{eq:polaLA4mmm}
\begin{align}
  \delta \varepsilon^{LA}_{x'x'} \propto \, & 3p_{12}+p_{c}+3(p_{13}-p_{12})\,,\\
  \delta \varepsilon^{LA}_{y'y'} \propto \, & 3p_{12}+p_{c}+\frac23(p_{33}-p_{11})+\frac13(p_{13}-p_{12})
  + \frac43(p_{31}-p_{12})+\frac83(p_{66}-p_{44})\,,
\end{align}
\end{subequations}
\begin{subequations}\label{eq:polaTAy4mmm}
\begin{align}
  \delta \varepsilon^{TA_{2}}_{x'x'} \propto \, & p_{c}-2(p_{13}-p_{12})\,,\\
  \delta \varepsilon^{TA_{2}}_{y'y'} \propto \, & -p_{c}-\frac13(p_{33}-p_{11})-\frac23(p_{13}-p_{12})
  + \frac43(p_{31}-p_{12})-\frac43(p_{66}-p_{44})\,,
\end{align}
\end{subequations}
In this case, the signal can have a rather complex shape. 
We note that the signal for the TA$_{2}$ phonons is linearly polarized if the crystal is almost isotropic normal to the $[001]$ axis which means that $\delta{\varepsilon_{x'x'}^{TA_{2}}}\approx 0$, while $\delta{\varepsilon_{y'y'}^{TA_{2}}}$ is large. A realistic scenario is the combination of the photoelastic tensor being isotropic in the $x'y'$ plane (which leads to $p_c=0$) and the same effect of strain along and normal to the $c$ axis on the dielectric constant (which means that $p_{12}\approx p_{13}$).

4. In the $4/m$ system, there are 10 independent photoelastic tensor components, in addition to the $4/mmm$ case, there are the non-zero $p_{45}$, $p_{16}$ and $p_{61}\neq 0$ components. In comparison to $4/m$, their main effect is rotation of the signal with respect to the crystallographic axes. 

\end{document}